\documentclass{aa}
\usepackage{epsfig}
\usepackage{aalongtable}
\usepackage{txfonts}

\newcommand{\teff}{T$_{\rm eff}$}

\newcommand{\nfe}{$\log$~n(Fe)}

\begin{document}

\title{Element abundances of unevolved stars in the open cluster M~67
\thanks{Based on observations collected at ESO-VLT, Paranal Observatory,
Chile, Programme numbers 65.L-0427, 68.D-0491, 69.D-0454}}

   \subtitle{ }

   \author{S. Randich\inst{1} \and P. Sestito\inst{2} \and F. Primas\inst{3} 
         \and R. Pallavicini\inst{4} \and L. Pasquini\inst{3}}

   \offprints{S. Randich, email:randich@arcetri.astro.it}

\institute{INAF/Osservatorio Astrofisico di Arcetri, Largo E. Fermi 5,
             I-50125 Firenze, Italy
\and
INAF/Osservatorio Astronomico di Bologna, Via C. Ranzani 1, I-40127 Bologna,
Italy
\and
European Southern Observatory, Garching bei M\"unchen, Germany
\and
INAF/Osservatorio Astronomico di Palermo, Piazza del
                  Parlamento 1, I-90134 Palermo, Italy}

\titlerunning{Chemical abundances in M~67}
\date{Received Date: Accepted Date}


\abstract
{The star-to-star scatter in lithium abundances observed among otherwise
similar stars in the solar-age open cluster M~67 is one of the most puzzling
results in the context of the so called ``lithium problem". Among other
explanations,
the hypothesis has been proposed that the dispersion in Li is due
to star-to-star differences in 
Fe or other element abundances which are predicted
to affect Li depletion.
}
{The primary goal of this study is the determination of the 
metallicity ([Fe/H]),
$\alpha$- and Fe-peak abundances in a sample of Li-poor and
Li-rich stars belonging to M~67, in order to test this hypothesis.
By comparison with previous studies,
the present investigation also allows us to check for intrinsic differences
in the abundances of evolved and unevolved cluster stars and to draw
more secure conclusions on the abundance pattern of this cluster.}
{We have carried out an analysis of high resolution UVES/VLT
spectra of eight unevolved
and two slightly evolved cluster members using MOOG and measured equivalent
widths. For all the stars we have determined [Fe/H] and element
abundances for O, Na, Mg, Al, Si, Ca, Ti, Cr and Ni.}
{We find an average metallicity [Fe/H]~$=0.03\pm0.01$, in very good
agreement with previous determinations. All the [X/Fe] abundance ratios
are very close to solar.
The star-to-star scatter in [Fe/H] and [X/Fe] ratios for all
elements, including oxygen, is lower than
0.05~dex, implying that the large dispersion in lithium 
among cluster stars is not due to differences in these element
abundances. We also find that, when using a homogeneous
scale, the abundance pattern of unevolved stars
in our sample is very similar to that of evolved stars, suggesting
that, at least in this cluster, RGB and clump stars 
have not undergone any chemical
processing. Finally, our results show that M~67 has a chemical
composition that is representative of the solar neighborhood.}
   {}

\titlerunning{Chemical abundances in M~67}
\authorrunning{S. Randich et al.}
\maketitle
\section{Introduction}

``Classical" or ``standard" models of stellar evolution include
convection only
as internal mixing process and neglect more complex physical processes
like diffusion, magnetic fields, rotation; these models 
predict that solar-type stars should not
deplete lithium (Li)  while on the main sequence (MS), since the 
base of the convective
zone does not reach the layers in the stellar interior where the temperature
is high enough for Li reactions.
Furthermore, according to standard models, stars with the same age, mass,
and chemical composition should undergo the same amount of Li depletion.
In sharp contrast with these predictions, not only is now well established
on observational grounds that
solar-type stars, including the Sun, do deplete a significant amount
of Li during the MS phases, but Li depletion can be different for
similar stars (e.g., Randich \cite{r_cast} and references therein).
The factor of about 10 star-to-star scatter in 
Li abundances observed among otherwise
identical F- and G-type members of the 4.5~Gyr old cluster M~67 indeed
represents one of the most puzzling results in the context of the
so-called ``Lithium problem"
(Spite et al. \cite{spi87}; Garcia L\'opez et al.
\cite{gar88}; Pasquini et al. \cite{pas97}; Jones et al. \cite {jon99}).

A large dispersion in Li abundances
is present among old field stars (e.g., Pallavicini et al. \cite{pal87};
Pasquini et al. \cite{pas94}). On the other hand,
whereas no dispersion is seen in the 600~Myr old Hyades,
in the three  2~Gyr old clusters IC~4651, NGC~3680
and NGC~752 (Randich et al. \cite {r00}; Sestito et al. \cite{ses_752}),
nor in 6~Gyr old NGC~188 (Randich et al \cite{n188}),
preliminary results of Li measurements among
large samples of stars in the 2~Gyr, metal rich NGC~6253
and in the very old Collinder~261 suggest that they could be
characterized by some amount of scatter (Pallavicini et al.
\cite{pall_cast}; Randich \cite{r_cast}). In other words,
the appearance of a dispersion seems to depend
more on the characteristics of the cluster than on age.

It has been shown that non membership and/or binarity are not
the reasons for the scatter in M~67 (Pasquini et al. \cite{pas97}); 
also, the effects
of chromospheric activity on the formation of the Li~{\sc i} line, which
are proposed as a possible explanation for the dispersion
in young clusters (Jeffries \cite{jef_06}
and references therein), are unlikely to play a role, given that at the old
age of M~67 the level of chromospheric activity should be low enough not to
affect the Li~{\sc i} line. Hence,
the scatter is most likely intrinsic and,
under the very reasonable assumption that cluster stars were all
born with the same Li content, it
must reflect different amounts of Li depletion. 

Different possibilities were proposed to explain the existence of
the spread (Randich \cite{r_cast}); among them,  
it has been suggested that
M~67 members do not have an identical chemical composition
and that the scatter in Li may reflect a scatter in iron content
or, more in general, in heavy element composition (Garcia
L\'opez et al. \cite{gar88}; Piau et al. \cite{piau03}).

As well known, chemical composition
affects stellar opacities and thus internal structure,  
mixing processes (both standard and non standard ones) and, in principle,
Li depletion. The effect of variations of the chemical composition
on pre-main sequence Li depletion has been theoretically investigated
in different studies (e.g, Swenson et al. \cite{sf_92}; Piau \& Turck-Chieze
\cite{ptc}; Sestito et al. \cite{ses_06}) and all of them agree in that
even relatively small changes of the mass fraction of elements critical
for the opacity can have significant effects on the amount of Li depletion. 
Similarly, Piau et al. (\cite{piau03}) have shown that 
variations in CNO abundances can change the amount of
Li depletion during the MS phases and suggested that star-to-star 
differences in these element abundances or, more in general
in $\alpha$ and Fe-peak abundances,
could explain the observed scatter in M~67. More quantitatively, they found
that a difference of 0.05~dex in [CNO/Fe] would
result in a difference in $\log$~n(Li) of $\sim 0.5$~dex (i.e, smaller
than the observed spread), implying that larger differences in CNO
abundances are needed to explain the whole spread.

Garcia L\'opez et al. (\cite{gar88}) 
did not find any difference in the overall metallicity of
Li-poor and Li-rich stars and ruled out that different [Fe/H] values within M~67
could be the reason for the scatter in Li. Here we extend their work to elements
other than iron and to a larger sample of stars, 
to investigate {\it a)} whether a large ($> 0.05$~dex)
star-to-star scatter in heavy elements
is present among unevolved cluster members; and, in case such a
spread is detected, {\it b)} whether it is related to the dispersion in Li.

In a more general context, the determination of the metallicity
and chemical composition of open clusters covering a large interval
of ages, metallicities, and Galactocentric distances is a critical tool
to investigate the formation and evolution of the Galactic disk
(Friel \cite{friel} and references therein). 
M~67 is one of the closest and best studied old open clusters; nevertheless,
relatively few studies focused on the determination of its chemical
composition (Tautvai\v sien\. e et al.  \cite{tau00} and references therein), 
and most of them
are based on few and/or evolved stars that may have undergone
chemical processing. The present work represents the first
abundance study of M~67 based on a 
rather large sample of unevolved (or slightly evolved) cluster members,
allowing us to check whether intrinsic 
differences exist between abundances of unevolved and evolved stars
and to draw more secure conclusions on the abundance pattern
of this cluster.
The paper is structured as follows:
in Sect.~2 the sample and the observations are described, while
the abundance analysis is presented in Sect.~3. The results, together
with a discussion of internal and systematic errors, and
a comparison with findings from previous studies are presented in Sect.~4.
Finally, a discussion of the results and conclusions are given in Sect.~5.
\section{Sample and observations}
\begin{figure*}
\psfig{figure=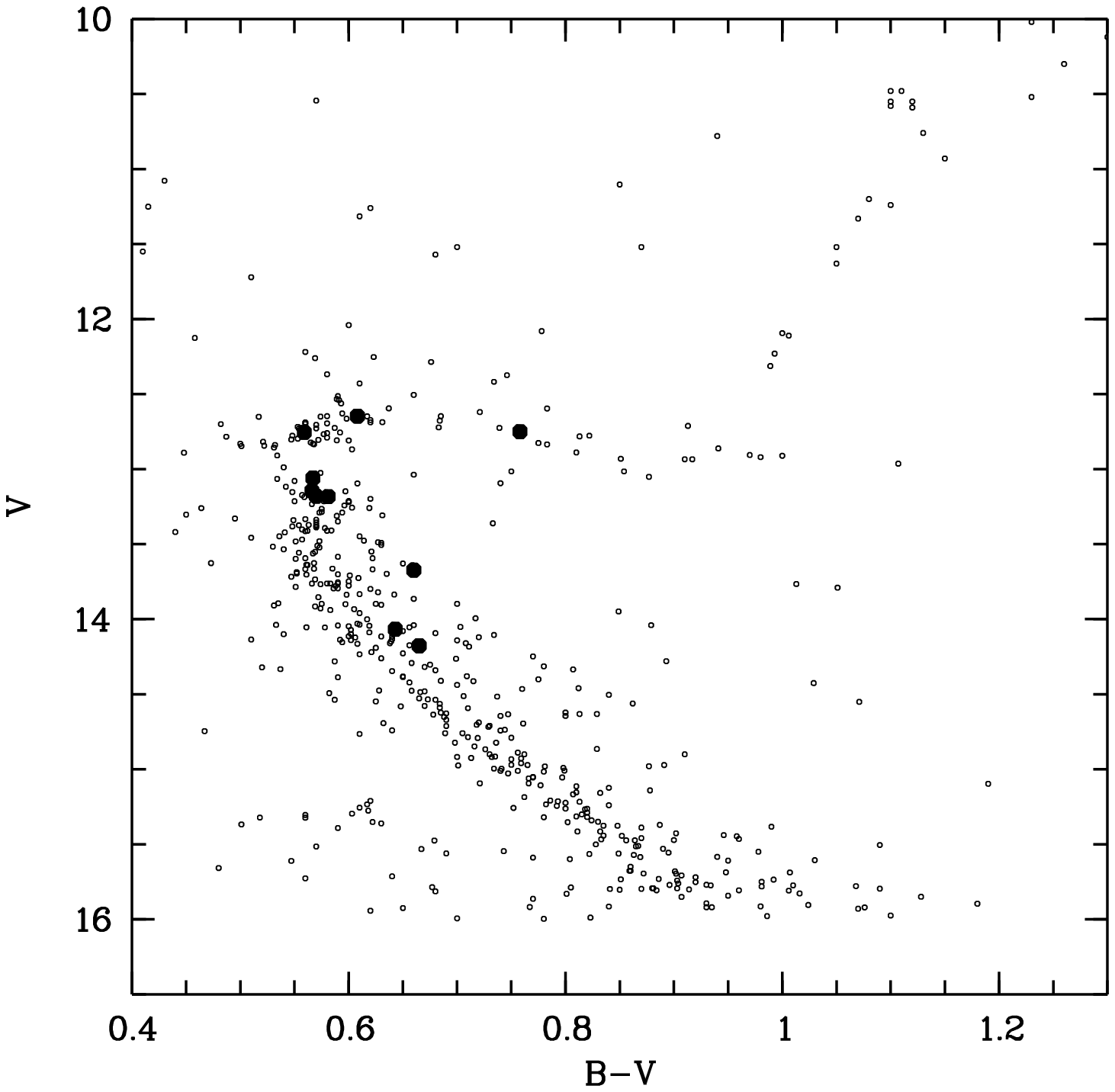, width=12cm}
\caption{Color-magnitude diagram of M~67. The sample stars are indicated
as filled circles. CCD photometry was taken from Montgomery et al.
(\cite{mont93}). Only stars brighter than V=16 are plotted in the
figure.} \label{fig_cm}
\end{figure*}

Our sample includes 10 single M~67 members that are 
listed in Table~\ref{sample}, 
together with V, $B-V$, and 
($B-V)_0$ colors. The color-magnitude diagram of
the cluster with our sample stars
evidenced in black is shown in Fig.~\ref{fig_cm}; the diagram
indicates that seven of them are still on the MS, one is
close to the the turn-off, while the remaining two are
already evolved to the subgiant branch. 

The observations were obtained using the UVES spectrograph (Dekker et al. 
\cite{dek}) on VLT UT2/Kueyen during three different
observing runs; the first one was carried out in Visitor mode in April 2000, 
while the other two were performed in Service mode in Fall 2001 and
Spring 2002, respectively.
Details on the observations, whose primary goal was the measurement 
of beryllium abundances in the sample stars,
were already given in Randich et al. (\cite{R02}) and we summarize 
the main points in the following. 
UVES was operated in Dichroic Mode using Cross Dispersers \#1 and \#3 in the
Blue and Red arms, respectively. 
The Red arm, which is of interest here, is equipped with a mosaic
of two CCDs composed by a 
EEV 2048$\times$4102 CCD and a MIT-LL $2048\times 4102$
CCD; the spectral coverage ranges
from approximately 4780 to 6810~\AA. The 15$\mu m$ pixels
together with a 1~arcsec wide
slit (projecting onto 4 pixels)
and CCD binning $1\times 1$, yielded a resolving power R$\sim 45,000$.
Exposure times were set based on the requirement of a good S/N
ratio in the near-UV spectral regions where the Be~{\sc ii} lines
are located and range between 1.3 and 3 hrs per star. 
Data reduction was carried out using the UVES pipeline and following the 
usual steps.
Typical $S/N$ ratios per resolution element
measured on the extracted 1-D spectra range between 90 and 180.
\setcounter{table}{0}
\begin{table*}
\caption{Sample stars, photometry,
stellar parameters and derived metallicities together with rms scatter.
Numbering is from Sanders (\cite{sand}).
V magnitudes and $B-V$ colors were taken from
Montgomery et al. (\cite{mont93}). ($B-V)_0$ colors were instead
retrieved from Jones et al. (\cite{jon99}) who had
derived them based on original $B-V$ and $V-I$ colors of Montgomery
et al. and assuming a reddening 
E($B-V)=0.05$. For stars S1034 and S1239,
not included in the sample of Jones et al. (\cite{jon99}),
we obtained ($B-V)_0$ colors in the same way. 
One Hyades member is also included (see text).}\label{sample} 
\begin{tabular}{rcccccrcc}
  & & & &  & & & & \\ \hline
 S & V & $B-V$ & ($B-V)_0$ & T$_{\rm eff}$ & $\xi$ & $\log$~g & [Fe/H]{\sc i} &rms \\
   &   & & & (K) & (km/s) & & \\
  & & & & &  & &  & \\
 969  & 14.18 & 0.665 & 0.622 & 5800 & 1.10 & 4.4 & 0.01 &  0.04  \\
 988  & 13.18 & 0.570 & 0.534 & 6153 & 1.45 & 4.1 & 0.03 &  0.04 \\
 994  & 13.18 & 0.581 & 0.535 & 6151 & 1.45 & 4.1 & 0.0  &  0.04   \\
 995  & 12.76 & 0.559 & 0.521 & 6210 & 1.50 & 3.9 & 0.05 &  0.03  \\
 998  & 13.06 & 0.567 & 0.518 & 6223 & 1.50 & 4.0 & 0.07 &  0.05  \\
 1034 & 12.65 & 0.608 & 0.567 & 6019 & 1.50 & 4.0 & 0.01 &  0.05   \\
 1239 & 12.75 & 0.758 & 0.692 & 5541 & 1.25 & 3.8 & 0.02 &  0.03  \\
 1252 & 14.07 & 0.643 & 0.588 & 5938 & 1.15 & 4.4 & 0.05 &  0.03  \\
 1256 & 13.67 & 0.660 & 0.595 & 5907 & 1.10 & 4.5 & 0.06 &  0.03 \\
 2205 & 13.14 & 0.566 & 0.534 & 6156 & 1.45 & 4.1 & 0.00 &  0.05 \\
  & & & & &  & &  & \\
vB 187 & 9.01  & 0.76  & 0.75  & 5339 & 0.9  & 4.5 & 0.13 & 0.05 \\ \hline
\end{tabular}
\end{table*}
\section{Abundance analysis}
Abundance analysis was carried out by means of measured equivalent widths
($EW$s) and using Version 2000 of 
MOOG (Sneden \cite{sne_moog}) with a grid of 1-D model atmospheres
from Kurucz (\cite{kur93}).
We recall that MOOG performs a standard-LTE analysis.
\subsection{Line list and equivalent widths}
Spectral lines to be used for the analysis 
of Fe~{\sc i}, Fe~{\sc ii} and, other elements (Na, Mg, Al, Si, Ca, Ti,
Cr, Ni)
were selected from different sources in the literature
and subsequently checked 
for suitability (in particular for blends) on the solar spectrum
obtained with UVES at the same resolution as our sample stars.
We finally retained in the list 55 Fe~{\sc i} and 11 Fe~{\sc ii}
lines, and a total of approximately 70 lines for all the other elements.

The task {\it eq} in SPECTRE was used to interactively measure the
$EW$s of the spectral lines by gaussian fitting.
Although the spectra had been previously normalized, local continuum 
was inspected and, if needed, adjusted at each $EW$ measurement.
\subsection{Atomic parameters and solar analysis}
The majority of $\log gf$ values for both Fe and other element lines were taken 
from the Vienna Atomic Line Data-base (VALD: Piskunov et al. \cite{pis95}; 
Kupka et al \cite{kup99}: Ryabchikova et al. \cite{rya99} ---
http://www.astro.uu.se/htbin/vald).
For a few lines for which $\log gf$ from VALD were
not available or gave very discrepant abundances for the Sun, $\log gf$
were either taken from other sources in the literature or, in a few cases,
adjusted through an inverse solar analysis. Note that
$\log gf$ values for all Fe~{\sc i}
lines were taken from VALD.
In Table~2 we show all the lines included in our
list together with their $\log gf$ values,
the source for these values, and the $EW$s measured in the solar
spectrum. 
Radiative and Stark broadening are treated in a standard way in MOOG;
as for collisional broadening, we used the Uns\"old approximation (\cite{uns})
for all the lines. As discussed by Paulson et al. (\cite{paul03}) this choice
should not greatly affect the differential analysis with respect
to the Sun.  We also mention that
very strong lines that are most affected by the treatment of damping
have been excluded from our analysis.

The analysis of the
solar spectrum was performed assuming the following solar 
parameters: T$_{\rm eff\;\odot}=5770$~K, $\log g_{\odot}=4.44$
and $\xi_{\odot}=1.1$~km/s. We mention that
$\log$ n(Fe~{\sc i}) vs. $EW$ did not show any trend, 
implying that the assumed
microturbulence is correct. On the contrary we found
a small, but statistically significant (slope equal to 0.009,
correlation coefficient 0.326) trend of Fe~{\sc i} abundances
vs. excitation potential (EP). Such a trend would disappear by 
assuming a 70~K higher solar effective temperature
(\teff$_{\odot}=5840$~K), which would result in larger and smaller
abundances for Fe~{\sc i} and Fe~{\sc ii}, respectively
($\log$~n(Fe~{\sc i})=7.58 and $\log$~n(Fe~{\sc ii})=7.49).

Output solar abundances are listed in Table~\ref{sun}
together with those from Anders \& Grevesse (\cite{ag89})
that are used as input in MOOG. Note that,
although solar abundances from Anders \& Grevesse (\cite{ag89})
have been superseded
by the study of Grevesse \& Sauval (\cite{grsa_98}), the two studies
yield the same abundances or very little differences for the elements
analyzed in this study.
Table~\ref{sun} shows a good 
agreement between the abundances determined
by us and the values of Anders \& Grevesse (\cite{ag89}), the most
discrepant element being Na with $\Delta \log$(Na)=0.05~dex.
We also find a very good agreement between $\log$~n(Fe) 
from Fe~{\sc i} and Fe~{\sc ii}.
\setcounter{table}{2}
\begin{table}
\caption{Solar abundances from Anders \& Grevesse (\cite{ag89})
and those determined through our analysis. Note that $\log gf$ values
for Al were derived by an inverse solar analysis.}\label{sun}
\begin{tabular}{ccc} \hline
 & & \\
Element & $\log$~n(X)$_{\rm AG89}$ & $\log$~n(X)$_{\rm our}$ \\
 & & \\ 
O~{\sc i}  & 8.87 & 8.66$\pm 0.04$\\ 
Na~{\sc i} & 6.33 & 6.28$\pm 0.03$\\
Mg~{\sc i} & 7.58 & 7.57$\pm 0.01$ \\
Al~{\sc i} & 6.47 & 6.47$\pm 0.03$ \\
Si~{\sc i} & 7.55 & 7.56$\pm 0.03$ \\
Ca~{\sc i} & 6.36 & 6.35$\pm 0.02$\\
Ti~{\sc i} & 4.99 & 4.97$\pm 0.02$\\
Cr~{\sc i} & 5.67 & 5.65$\pm 0.02$\\
Fe~{\sc i} & 7.52 & 7.52$\pm 0.03$\\
Fe~{\sc ii} & 7.52 & 7.51 $\pm 0.04$\\
Ni~{\sc i} & 6.25 & 6.25 $\pm 0.03$ \\  \hline
\end{tabular}
\end{table}
\subsection{Oxygen}
We do not discuss here all the intricacies related to oxygen
abundance determinations in stars, and we
refer to Bensby et al. (\cite{bensby04})
and Schuler et al. (\cite{schuler}) for recent discussions. The forbidden
lines at 6300.30~\AA~and 6363.78~\AA~are not significantly sensitive to NLTE 
effects nor to stellar
effective temperature; therefore, as widely acknowledged, these lines are
the most reliable ones for deriving O abundances. 
Our O determinations are based on the [O~{\sc i}] forbidden line at
6300.3~\AA. Although this line is located in a spectral region that is
severely affected by the presence of telluric lines, the radial
velocity of M~67 is such that they do not affect the [O~{\sc i}] line in
our spectra. 

As first pointed out by Lambert (\cite{lambert}), the 6300.3~\AA~
line is blended with a Ni~{\sc i} feature whose contribution
must be taken properly into account for an accurate determination of
O abundance. Hence,
we determined O both for the Sun and our
sample stars using the driver {\it blend} in MOOG, that allows accounting for
the blending Ni~{\sc i} 6300.34 feature when force fitting the abundance of the
O to the measured $EW$s of the 6300.3~\AA~feature.
As done for the other lines, $EW$s of the 6300.3~\AA~feature
were measured by gaussian fitting using SPECTRE. The
solar [O~{\sc i}] $EW$ measured in the UVES spectrum is 5.5 $\pm 0.3$~m\AA,
in good agreement with previous estimates (see Schuler et al \cite{schuler});
$EW$s for the sample stars 
are listed in the second column of Table~\ref{oxygen}.

For our analysis
we employed the [O~{\sc i}] $gf$-value determined by Allende Prieto et al. 
(\cite{ap01}), $\log~gf=-9.717$, while for the Ni~{\sc i} 6300.34~\AA~blend,
following Johansson et al. (\cite{joha03}), we used $\log~gf=-2.11$.
Note that, as discussed by Schuler et al., we would have obtained
virtually the same O abundances by considering the two separate components
of the Ni~{\sc i} blend.
Assuming a solar Ni abundance $\log$~n(Ni)=6.25 (see Table~\ref{sun}),
we obtained for the Sun $\log$~n(O)=8.66~$\pm 0.04$, much below
the classical value of Anders \& Grevesse (\cite{ag89}), but in 
good agreement
with recent determinations, including those obtained using 3-D
analysis (see again the discussion in Schuler et al.).
\subsection{Stellar parameters}
Initial stellar parameters for the sample stars were 
estimated from photometry (see Table~\ref{sample}).
Effective temperatures were retrieved from
Jones et al. (\cite{jon99}), who, in turn, had derived them 
using the calibration of Soderblom et al. (\cite{sod93}), i.e, \teff
$=1808\times (\rm B-\rm V)_0^2-6103\times (B-V)_0+8899$.
Microturbulence velocities were estimated as: $\xi=3.2\times 10^{-4}
(\rm T_{\rm eff}-6390)-1.3(log g-4.16) + 1.7$
(Nissen \cite{nis81}), while surface
gravities were derived using the relationship:
$\log g=4.44 + \log \rm M/\rm M_{\odot}+4 \log \rm T_{\rm eff}/\rm T_{\rm eff}\;
{\odot}+0.4(\rm M_{\rm bol}-\rm M_{\rm bol}\;{\odot}$). Absolute bolometric
magnitudes were determined from V magnitudes, adopting
a distance modulus (m--M)$_0$=9.6 (Pace et al. \cite{pace_04}:
Sandquist \cite{san04}) and assuming
bolometric corrections from Johnson (\cite{john_66}). Finally, stellar masses
were derived following Balachandran~(\cite{bala95}).

In order to determine spectroscopic temperatures one would need 
to change the 
initial \teff~values until no \nfe{\sc i} vs. EP trend is seen. However,
we decided not to change initial \teff~derived from photometry,
since a trend is present for the solar spectrum and
for all the sample stars we found similar slopes (in magnitude and sign).
To further test
the correctness of our approach, for each star we plotted $\Delta$\nfe$=$
\nfe$_{\odot}-$\nfe$_{\rm star}$ as a function of EP
and checked that no significant residual trends
were present. Spectroscopic microturbulence values for each star were instead
obtained by changing the input microturbulence 
until no \nfe~vs. $EW$ trends were seen.
Finally, $\log$~g values were varied only when necessary, in order to have
a difference between iron abundances derived from Fe~{\sc i} and Fe~{\sc ii}
lines not greater than 0.05~dex, i.e., consistent, within errors,
with the ionization equilibrium. It turned out that we needed to
change the initial values of gravity for two stars only, S1239 and S1256.
Assumed parameters for the sample stars
are listed in Cols. 5--7 in Table~\ref{sample}.
\section{Results}
\subsection{Abundances}
Log~n(X) values for each line were determined based on measured $EW$s
and stellar parameters listed in Table~\ref{sample}. 
Final abundances for each star and each element were determined as the
mean abundance from the different lines. 
1$\sigma$ clipping was performed
for iron, but not for the other elements for which a smaller number
of lines was available. 
As to oxygen, [O/H] and [O/Fe] values for the sample stars were 
obtained from the measured $EW$ of the 6300.3~\AA~blend (see Table~\ref{oxygen})
and assuming Ni abundances determined through our
analysis.
As already mentioned, [Fe/H] and [X/Fe] ratios
for each star were determined differentially with respect to solar
abundances listed in Table~\ref{sun}.
 
[Fe/H]{\sc i}
abundances and rms values for the sample stars are listed in
Cols.~8 and 9 of Table~\ref{sample}, while 
in Fig.~\ref{figure2} we show [Fe/H] as a function of effective temperature.
The figure clearly shows that no trend of [Fe/H] as a function of
\teff~is present. 
\setcounter{table}{3}
\begin{table*}
\caption{Derived abundance ratios for M~67 stars and
one Hyades member. For each star,
quoted errors on abundance ratios are the quadratic sum
of rms values of [Fe/H] and [X/H]. Average abundance ratios for the Hyades,
taken from Friel (\cite{friel}), are given in the last line.} \label{ratios}
\scriptsize
\begin{tabular}{lrrrrrrrr}\hline
 S & [Na/Fe] & [Mg/Fe] & [Al/Fe] & [Si/Fe] & [Ca/Fe] & [Ti/Fe] & [Cr/Fe] & [Ni/Fe] \\
   &  & & & & & &  &\\ 
 969 &  $0.00\pm 0.06$ & $-0.01\pm 0.05$ & $-0.01\pm 0.07$ & $-0.01\pm 0.06$ & $0.09\pm 0.06$ & $0.00\pm 0.07$ & $0.00\pm 0.06$ & $-0.04\pm 0.07$ \\
  988 &  $0.12\pm 0.04$ & $0.03\pm 0.13$ &$-0.04\pm 0.06$ & $0.10\pm 0.05$ & $0.12\pm 0.06$ & $0.00\pm 0.06$ & $0.09\pm 0.04$ & $0.03\pm 0.06$ \\
  994 &  $0.12\pm 0.05$ & $0.02\pm 0.09$ & $-0.04\pm 0.04$ & $0.02\pm 0.07$ & $0.07\pm 0.06$ & $0.04\pm 0.06$ & $0.03\pm 0.05$ & $0.04\pm 0.06$ \\
  995 & $0.04\pm 0.04$ & $0.00\pm 0.13$ & $-0.06\pm 0.08$ & $0.03\pm 0.06$ & $0.05\pm 0.05$ &$-0.06\pm 0.08$ & $-0.04\pm 0.04$ & $-0.07\pm 0.04$ \\
  998 &  $0.08\pm 0.06$ & $0.00\pm 0.09$ &$-0.08\pm 0.07$ & $-0.02\pm 0.07$ & $0.01\pm 0.07$ & $0.00\pm 0.08$ & $-0.05\pm 0.05$ & $-0.03\pm 0.06$ \\
 1034 &  $0.07\pm 0.06$ & $-$0.02$\pm 0.05$ &$-0.08\pm 0.04$ & $0.03\pm 0.06$ & $0.03\pm 0.06$ & $0.00\pm 0.07$ &  $0.00\pm 0.05$ & $-0.03\pm 0.07$ \\
 1239 & $0.03\pm 0.09$ & $-$0.01$\pm 0.09$ & $0.02\pm 0.06$ & $0.05\pm 0.07$ & $0.00\pm 0.07$ & $-0.05\pm 0.05$ & $-0.01\pm 0.05$ & $-0.03\pm 0.05$ \\
 1252 & $-0.04\pm 0.06$ & $-0.04\pm 0.03$ & $-0.09\pm 0.06$ & $0.00\pm 0.05$ & $0.02\pm 0.04$ & $-0.08\pm 0.06$ & $-0.04\pm 0.04$ & $-0.05\pm 0.05$ \\
 1256 & $-0.06\pm 0.04$ & $-0.04\pm 0.04$ & $-0.11\pm 0.05$ & $-0.03\pm 0.05$ & $0.04\pm 0.06$ &$-0.04\pm 0.05$ & $-0.02\pm 0.07$ & $-0.04\pm 0.05$ \\
 2205 &  $0.13\pm 0.06$ &  $0.03\pm 0.07$ & $-0.06\pm 0.08$ & $0.04\pm 0.08$ & $+0.03\pm 0.06$ & $-0.06\pm 0.09$ & $-0.01\pm 0.05$ & $0.03\pm 0.06$ \\
   &  & & & & & &  &\\ 
vB187 & $+0.00 \pm 0.05$ & $-0.06\pm 0.02$ & $-0.07\pm 0.07$ &0.07 $\pm 0.07$ & $0.04\pm 0.06$ & $-0.09 \pm 0.04$ & $-0.05\pm 0.02$ & $0.03\pm 0.06$ \\
Hyades$_{\rm lit.}$ & $0.01$ & $-0.06$ & $-0.05$ & $0.04$ & $+0.06$ & $-0.06$ & --- & --- \\ \hline
\end{tabular}
\end{table*}
\setcounter{table}{4}
\begin{table*}
\caption{[O~{\sc i}] 6300~\AA~line $EW$s, [O/H] and [O/Fe] values
for the sample stars and the Hyades member. 
Quoted errors in [O/H] are due to uncertainties
in $EW$s, while errors in [O/Fe] are the quadratic sums of errors in [O/H] and
rms values of [Fe/H] (see Table~\ref{sample}). In the last line we provide
average [O/H] and [O/Fe] ratios for the Hyades from Schuler et 
al. (\cite{schuler}).}\label{oxygen}
\begin{tabular}{lcrr} \hline
 S & $EW$ (6300.3~\AA) & [O/H] & [O/Fe] \\
  & (m\AA) &  &  \\
  & & & \\
  969 &  5.6 $\pm 0.5$    &  0.02 $\pm 0.06$ & 0.01 $\pm 0.07$\\
  988 &  5.3 $\pm 0.9$    &  0.00 $\pm 0.10$ & $-0.03\pm 0.11$\\
  994 &  5.0 $\pm 0.9$    &  $-0.03 \pm 0.10$ & $-0.03 \pm 0.11$\\
  995 &  6.6 $\pm 0.9$    & $+0.08 \pm 0.08$ & $0.03 \pm 0.08$\\
  998 &  5.7 $\pm 0.5$    & $+0.04 \pm 0.04$ & $-0.03 \pm 0.06$\\
 1034 &  6.5 $\pm 0.5$    & $+0.03 \pm 0.04$ & $+0.02 \pm 0.06$\\
 1239 & 10.2 $\pm 1$      & $+0.03 \pm 0.06$ & $+0.01 \pm 0.07$\\
 1252 &  5.1 $\pm 0.4$    & $+0.00 \pm 0.05$ & $-0.05 \pm 0.06$\\
 1256 &  5.0 $\pm 0.5$    & $+0.03 \pm 0.05$ & $-0.03 \pm 0.07$\\
 2205 &  5.4 $\pm 1$      & $+0.01 \pm 0.11$ & $0.01 \pm 0.12$\\
   &  & \\
 vB187 &  5.6 $\pm 0.4$   & $-0.19 \pm 0.08$ & $-0.32 \pm 0.09$\\
Hyades$_{\rm Schuler}$ & & 0.19 & 0.10 \\ \hline
\end{tabular}
\end{table*}
Using all the stars in our sample,
we obtain a mean value [Fe/H]$=0.03 \pm 0.01$ (rms$=0.03$, 10 stars),
very close to solar. This mean value is listed in the first column of
Table~\ref{taut_c}.

The [X/Fe] ratios for Na, Mg, Al, Si, Ca, Ti, Cr and Ni
are listed in Table~\ref{ratios}.
Errors on [X/Fe] values
in Table~\ref{ratios} correspond to 
the quadratic sum of rms of [Fe/H] and [X/H] values.
Results for O are listed separately in 
Table~\ref{oxygen}, where both [O/H] and [O/Fe] values are given; 
in this case errors in [O/H] correspond to uncertainties in
the measured $EW$s of the forbidden line, while errors in [O/Fe]
are the quadratic sum of errors in [O/H] and rms of [Fe/H].
Mean abundance ratios for M~67 together with 1$\sigma$ standard deviation
are listed in the second column of Table~\ref{taut_c}, while
[X/Fe] ratios for all elements are plotted in Fig.~\ref{figure3}
as a function of effective temperature. 
The figure shows that, as in the case of [Fe/H], no evident trends
of [X/Fe] ratios with \teff~are present. 
A small amount of star-to-star scatter
might be present, both in [Fe/H] and [X/Fe] ratios, but for all the
elements the scatter is well within measurement 
uncertainties. 
A more detailed discussion on the possible presence of a scatter
will be provided in Sect.~\ref{scatter}.
\subsection{Errors}
Sources of internal errors include uncertainties in atomic
and stellar parameters, as well as errors in measurements of $EW$s.
The sample spectra are 
characterized by different S/N ratios and it
is not possible to estimate a typical error in $EW$s; 
however, errors in the derived abundances due to 
errors in $EW$s are in a good
approximation represented by the standard deviation (or rms) 
around the mean abundance determined from individual lines. 
The rms in principle includes also errors due to uncertainties in atomic
parameters, but the latter should be minimized in our analysis,
since it is carried out differentially with respect to the Sun
and our stars have parameters (\teff, $\xi$, $\log$~g) close 
to the solar ones.  As already mentioned, rms values
for [Fe/H] are listed in Table~\ref{sample},
while errors in [X/Fe] ratios 
listed in Table~\ref{ratios} correspond to the quadratic
sum of rms for [Fe/H] and rms for [X/H]. 

Errors in [O/H] include in principle uncertainties in the measurement
of the $EW$ of the [O~{\sc i}]~6300.3~\AA~feature and errors due to
uncertainties in Ni abundance, which affect the estimate of the
contribution of the Ni~{\sc I} 6300.34 blend to the [O~{\sc i}] feature.
We find however that the latter are much smaller than typical errors
in [O~{\sc i}] $EW$s. Our spectra are not characterized by an extremely high
S/N in the forbidden line region and thus errors in $EW$s are significant.
Typical errors are of the order of 0.5-1~m\AA, which reflect into
uncertainties in [O/H] between 0.05 and 0.1~dex. On the other hand, 
uncertainties in Ni abundance are of the order of 0.05~dex at most and
correspond to errors in the Ni~{\sc i} $EW$ of the order of 0.15-0.2~m\AA.

Internal errors due to uncertainties in stellar parameters
were estimated
by varying each parameter separately, while leaving the other two
unchanged. We assumed random uncertainties of $\pm 70$~K, 
$\pm 0.15$~km/s,
and $\pm 0.25$~dex in \teff, $\xi$ and $\log$~g, respectively.
We performed different tests and found that for 
all the stars, \teff~variations larger than 70~K
would have introduced significant (i.e,
much larger than in the Sun) trends of $\log$~n(Fe~{\sc i})~vs. EP, while
variations in $\xi$ larger than 0.15 km/s would have resulted
in significant trends of $\log$~n(Fe) vs. $EW$. This applied not only to Fe,
but also to the other elements with several lines
covering large dynamical ranges in EP and $EW$s.
Finally, differences in $log$~g larger than 0.25~dex would have resulted
in differences between $\log$~n(Fe~{\sc i}) and $\log$~n(Fe~{\sc ii})
larger than 0.05~dex. 
In Table~\ref{errors} we list errors in [Fe/H] and [X/Fe] ratios due
to uncertainties in stellar parameters 
for the coolest (S1239) and one of the warmest (S988) stars
in the sample.

Systematic errors are more difficult to evaluate. Errors
in the scale of \teff~are likely small, since, as already noted,
we did not find major trends of inferred Fe abundances
as a function of EP.  By using the \teff~vs. $B-V$ calibration
of Alonso et al. (\cite{al96}), we would have obtained slightly
cooler temperatures, with differences ranging between 
40 and 60~K, that would have given a mean metallicity [Fe/H]=$-0.01$
(i.e., 0.04~dex below our estimate)
and almost identical [X/Fe] values. 
More in general, in order to get an estimate of
global systematic errors,
we analyzed the spectrum of one Hyades
member (vB187) observed with the same UVES set-up as our sample stars
(complete results on the chemical analysis of other Hyades members will be
reported elsewhere). 
The results for [Fe/H] and [X/Fe] for the Hyades member
are listed in Tables~\ref{sample}, \ref{ratios}, and \ref{oxygen};
our determination of the metallicity for this star is in perfect agreement
with the classical value for the Hyades 
(Boesgaard \& Friel~\cite{bf90}; Paulson et al. \cite{paul03}; 
Friel~\cite{friel}); 
as for the other elements, relatively few abundance determinations have been
carried out in the past, in spite of the fact that the Hyades is one of the
best studied open clusters. In the last line of Tables~\ref{ratios} and
\ref{oxygen}
we list the average [X/Fe] ratios for the Hyades taken from the
compilation of Friel~(\cite{friel} ---Table 1 in that paper). 
Our [X/Fe] ratios for most elements are in good agreement with
the values from the literature,
suggesting that our abundance scale should not be affected by major
systematic errors. 
On the other hand, for vB187 we derive an O
abundance (and [O/Fe] ratio) considerably below the average 
value of Schuler et al. (\cite{schuler}). We note however that the
discrepancy is most likely due to differences in the $EW$s of the
forbidden line, rather than to the analysis. For vB187 we measure an $EW$
of the [O~{\sc i}]~6300.3~\AA~line
of 5~m\AA~to be compared with values of 7-8~m\AA~found by Schuler et al.
for stars
with similar temperatures as vB187. Assuming their $EW$s, we would have
inferred an [O/H]  similar to their values.
\setcounter{table}{5}
\begin{table}
\caption{Random errors due to uncertainties in stellar parameters.}\label{errors}
\begin{tabular}{lccc}\hline
\multicolumn{4}{c}{S1239: \teff=5477~K, $\log$~g=3.6, $\xi=1.2$~km/s}\\
 & & &\\
$\Delta$ &  $\Delta$\teff$=\pm 70$ & $\Delta \log$g=$\pm 0.25$ & $\Delta \xi=\pm0.15$\\
   & (K)  & dex  & km/s \\
   &  &  &  \\
~[Fe/H]{\sc i}  & 0.06/$-0.04$   &  $-0.01$/0.02   & $-0.05$/0.06 \\
~[O/Fe]  & $-$0.05/0.03    &  0.12/$-$0.14   & $0.05$/$-0.06$ \\
~[Na/Fe] & $-$0.01/0.00    &  $-0.03$/0.01   & $0.03$/$-0.03$ \\
~[Mg/Fe] & $-$0.01/0.00    &  $-0.06$/0.05   & $0.02$/$-0.03$ \\
~[Al/Fe] & $-$0.02/0.00   &  $0.00$/$-$0.01 & $0.04$/$-0.05$ \\
~[Si/Fe] & $-$0.05/0.04    &  0.03/$-0.03$   & $0.03$/$-0.04$ \\
~[Ca/Fe] & 0.0/$-0.02$     &  $-0.04$/0.01   & $-0.01$/0.01 \\
~[Ti/Fe] & 0.02/$-0.04$ &  $0.00$/$-$0.01 & $0.01$/$-0.02$ \\
~[Cr/Fe] & 0.03/$-$0.03     &  $-0.01$/0.0    & $-0.02$/0.02 \\
~[Ni/Fe] & $-$0.01/0.01   &  0.02/$-0.03$   & $0.01$/$-0.01$ \\ \hline
 & & &\\
\multicolumn{4}{c}{S988: \teff=6151~K, $\log$~g=4.1, $\xi=1.45$~km/s}\\
 & & &\\
$\Delta$ &  $\Delta$\teff$=\pm 70$ & $\Delta \log$g=$\pm 0.25$ & $\Delta \xi=\pm0.15$\\
   & (K)  & dex  & km/s \\
   &  &  &  \\
~[Fe/H{\sc i}] & 0.05/$-0.05$  &  $-0.02$/0.02   & $-0.03$/0.04 \\
~[O/Fe]  & $-$0.04/0.04    &  0.13/$-$0.14   & $0.03$/$-0.04$ \\
~[Na/Fe] & $-$0.02/0.01 &  $-0.02$/0.01   & $0.01$/$-$0.03 \\
~[Mg/Fe] & $-$0.01/0.01  &  $-0.05$/0.05   & $0.02$/$-$0.03 \\
~[Al/Fe] & $-$0.02/0.02  &  $0.01$/$-$0.01 & $0.02$/$-$0.04 \\
~[Si/Fe] & $-$0.02/0.03  &  0.02/$-0.01$   & $0.02$/$-$0.03 \\
~[Ca/Fe] & 0.00/0.00     &  $-0.02$/0.02   & $-0.01$/0.01 \\
~[Ti/Fe] & 0.01/$-0.01$ &  $0.01$/$-$0.01 & $0.02$/$-$0.02 \\
~[Cr/Fe] & 0.01/$-0.02$ &  $0.00$/$-$0.01 & $-0.02$/0.0 \\
~[Ni/Fe] & $-$0.01/0.0   &  $0.01$/$-$0.02 & $0.01$/$-$0.02 \\ \hline
\end{tabular}
\end{table}
\subsection{NLTE effects and 3-D}
As well known, the assumption of LTE may introduce systematic errors
and may give origin to spurious
abundance trends when analyzing stars covering
large intervals of effective temperatures, gravities and metallicities.
NLTE effects depend on stellar temperature and gravity and
should not be a major concern in the present study,
since, as already stressed, we carried out a differential analysis
with respect to the Sun and most of our sample stars are similar to the Sun.
In any case,
Thevenin \& Idiart (\cite{thev}) computed NLTE corrections for Fe~{\sc i}
and Fe~{\sc ii} and showed that NLTE corrections are small for stars
with solar metallicity or higher.
At the metallicity of M~67 the lines used for Na determination are
marginally affected by NLTE effects (e.g., Mashonkina \cite{mash}):
in the temperature and gravity range of our sample stars
NLTE negative corrections
are always below 0.1~dex and differential corrections with respect
to the Sun are of the order
of 0.02--0.03~dex at most. NLTE corrections are also small
for the two Mg lines used in this study (Zhao et al.~\cite{zhao})
and the same holds for Al (Baum\"uller et al. \cite{bau}).

Use of time dependent, 3-D hydrodynamical model of the solar atmosphere
has resulted in the revision of solar abundances for several elements (Asplund
et al. \cite{asplund}). As for NLTE effects, use of 1-D models, should
not be a major concern for our differential analysis.
\subsection{Comparison with other studies}\label{sect_comp}
Our mean [Fe/H] value for M~67 is in very good  agreement with
the metallicity determinations of 
Garcia L\'opez et al. (\cite{gar88}, [Fe/H]=$0.04 \pm 0.04$),
Friel \& Boesgaard (\cite{fb92}, [Fe/H]=$0.02 \pm 0.12$) and
Yong et al. (\cite{yong05}, [Fe/H]$=0.02 \pm 0.14$),
while somewhat
higher than the values of
Tautvai\v sien\. e et al. (\cite{tau00}, [Fe/H]$=-0.03 \pm 0.03$) and
Shetrone \& Sandquist (\cite{shs00}, [Fe/H]$=-0.05$), but still
consistent with them.
The metallicity derived by Friel et al.
(\cite{friel02}, [Fe/H]=$-0.15 \pm 0.05$) from low resolution
spectroscopy of evolved cluster stars is instead far below our estimate.

To our knowledge, very few studies have been published
on the abundance pattern of M~67
and significant discrepancies exist between them.
Peterson (\cite{peters92}) reports a 0.1~dex $\alpha$-enhancement based 
on O, Mg, 
and Si measurements in two cluster giants; on the contrary,
Garcia L\'opez et al. 
(\cite{gar88}), based on the analysis of warm unevolved cluster stars,
find average Ca and Si abundances below solar ([Ca/H]$=-0.10 \pm 0.08$ or
[Ca/Fe]$=-0.14$,
[Si/H]=$-0.20 \pm 0.06$ or [Si/Fe]=$-0.24$). 
Both the Ca and Si analyses relied
on one line only. We have three stars in common with the sample
of Garcia L\'opez et al. (\cite{gar88}), namely, S988 (F129), S995 (F127),
and S2205 (F128); the agreement between their and our [Fe/H]
values is good, while we find higher Ca (significantly
higher in the case of F129) and much higher
Si abundances.

Tautvai\v sien\. e et al. (\cite{tau00}) carried out
a detailed chemical analysis of 9 clump and red giant branch (RGB)
cluster members, finding a rather normal (i.e., close to solar)
abundance pattern, apart from Na that appeared enhanced.
Similar results have been obtained more recently by 
Yong et al. (\cite{yong05}) based on three
clump cluster members (their stars are in common with the sample of
Tautvai\v sien\. e et al.), although their [X/Fe] ratios are slightly
higher than those of Tautvai\v sien\. e et al. In Table~\ref{taut_c} 
we compare our average [X/Fe] ratios to those of Tautvai\v sien\. e et al. 
for the elements determined in both studies. Specifically, we list
in Col.~2 the mean abundance ratios from the present study and in
Cols.~3 and 4 the average ratios from the whole sample of
Tautvai\v sien\. e et al.
and those obtained considering only the stars observed by them with a resolution
R=60,000. Focusing on the latter,
the table shows
that their [X/Fe] ratios are in general larger than ours
but, considering errors,
the abundance ratios for most elements
are consistent with each others. Also their [Fe/H] is
0.06~dex below our value, implying that the [X/H] ratios
are in better agreement.
The only exceptions are Na, Al, and Si, for which
they derive substantially higher [X/Fe] (or [X/H]) values than us.

These discrepancies could be either real, thus indicating 
intrinsic differences in the chemical composition of unevolved and evolved
cluster stars and hence chemical processing, 
or due to systematic offsets between Tautvai\v sien\. e et
al. (\cite{tau00})
and our abundance scale. In order to investigate this point,
we re-determined Fe, Na, Al, and Si abundances
for the three clump stars in their sample observed at high resolution,
using the same code and atomic parameters employed for our sample
stars, but their stellar parameters. For the analysis, we considered
only lines in their line list which were also included in ours.
In Table~\ref{taut_n} we compare the [X/Fe] derived by Tautvai\v sien\. e 
et al. (\cite{tau00})
with those obtained with our new analysis of these stars: 
the table shows that, for the
three elements and for the three stars, 
we find smaller [X/Fe] ratios than Tautvai\v sien\. e et al.,
with differences up to $\sim 0.15$~dex. Correspondingly,
the average ratios abundances of Na, Al, and Si to Fe determined through 
our reanalysis are lower than those
listed in the last column of Table~\ref{taut_c} and are all now consistent
with the mean ratios that we derive for unevolved stars. Sodium
remains slightly enhanced, but the difference between dwarfs and giants
can be explained by NLTE effects that are larger for cool giants than
for warm dwarfs (Mashonkina et al. \cite{mash}). 
In other words, our analysis suggests that no intrinsic differences are present
between unevolved and evolved stars in M~67, implying that the latter
have not undergone significant chemical processing.

To our knowledge, this is the first case where an agreement
between abundances in unevolved and evolved stars in the same open cluster
is found based on a significant sample of stars. In particular,
the good agreement of Na abundances for dwarf and giant stars
in the cluster suggests that the slight enhancement with
respect to the Sun is probably intrinsic to the cluster and not
due to the deep mixing in giants as hypothesized by Tautvai\v sien\. e 
et al. (\cite{tau00}) and by Pasquini et al. (\cite{pas04})
for IC~4651.

In summary, the comparison of our abundances for M~67 with those determined
by others shows
that systematic offsets exist between different studies; most obviously,
when investigating the abundance trends of open clusters as a function
of age or Galactocentric distances, one should  make sure that the data
are on the same abundance scale.
\setcounter{table}{6}
\begin{table}
\caption{Mean abundance ratios from this study and from 
Tautvai\v sien\. e et al. (\cite{tau00}).}\label{taut_c}
\begin{tabular}{llrcc}
\hline
  & Element& Present & Tautvai\v sien\. e & Tautvai\v sien\. e \\
  & ratio  & study  & all           & R=60,000  \\
  & & & &\\
 & [Fe/H]  & 0.03 $\pm 0.03$   & $-$0.03 $\pm 0.03$ & $-$0.02 $\pm 0.03$ \\
 & [O/Fe]  & 0.01 $\pm 0.03$   & 0.02 $\pm 0.06$ & 0.05$\pm 0.03$ \\
 & [Na/Fe] & 0.05 $\pm 0.07$   & 0.19 $\pm 0.06$ & 0.20 $\pm 0.01$ \\
 & [Mg/Fe] & 0.00  $\pm 0.02$   & 0.10 $\pm 0.04$ & 0.05 $\pm 0.02$ \\
 & [Al/Fe] & $-0.05 \pm 0.04$  & 0.14 $\pm 0.04$ & 0.10 $\pm 0.02$ \\
 & [Si/Fe] & $0.02 \pm 0.04$  & 0.10 $\pm 0.05$ & 0.08 $\pm 0.02$ \\
 & [Ca/Fe] & 0.05 $\pm 0.04$   & 0.04 $\pm 0.05$ & 0.00 $\pm 0.05$ \\
 & [Ti/Fe] & $-0.02 \pm 0.04$  & 0.04 $\pm 0.10$ & 0.03 $\pm 0.02$ \\
 & [Cr/Fe] & $-0.01 \pm 0.04$  & 0.10 $\pm 0.07$ & 0.04 $\pm 0.02$ \\
 & [Ni/Fe] & $-0.02 \pm 0.04$  & 0.04 $\pm 0.04$ & 0.01 $\pm 0.02$ \\ \hline
\end{tabular}
\label{giants}
\end{table}
\section{Discussion and conclusions}
\subsection{Star-to-star scatter and Li} \label{scatter}
As mentioned in Sect.~1,
the investigation of the presence (or lack thereof) of a significant
dispersion in $\alpha$ and Fe-peak element
abundances among cluster stars, that could possibly
explain the large star-to-star scatter in Li abundances, was the main
motivation for the present study. 

\begin{figure*}
\psfig{figure=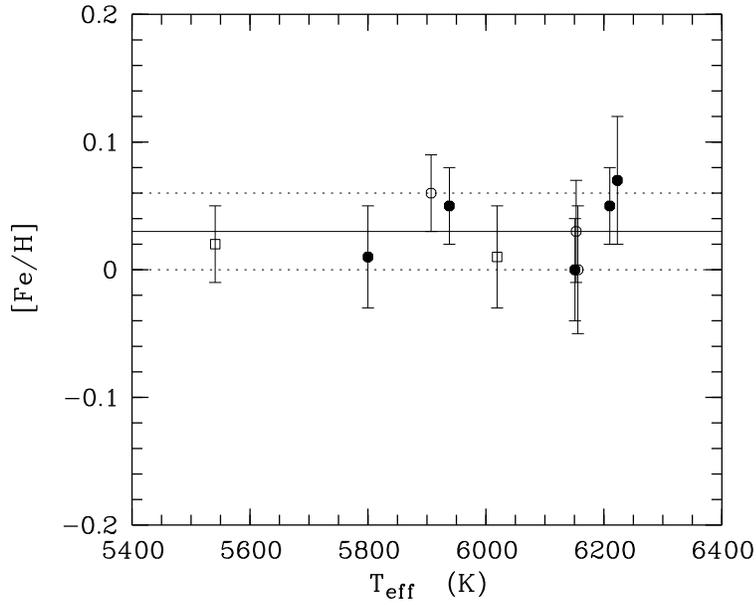, width=10cm,angle=-90}
\caption{[Fe/H] as a function of effective temperature for the sample
stars. Filled and open circles represent main sequence 
Li-rich and Li-poor stars, while open squares denote the two slightly
evolved stars.
Error bars shown in the figure correspond to rms 
values quoted in Table~\ref{sample}. 
Internal errors
due to uncertainties in stellar parameters are not shown here.
The solid
and dotted lines represent the average [Fe/H] value $\pm 1$ standard
deviation.}\label{figure2}
\end{figure*}

\begin{figure*}
\psfig{figure=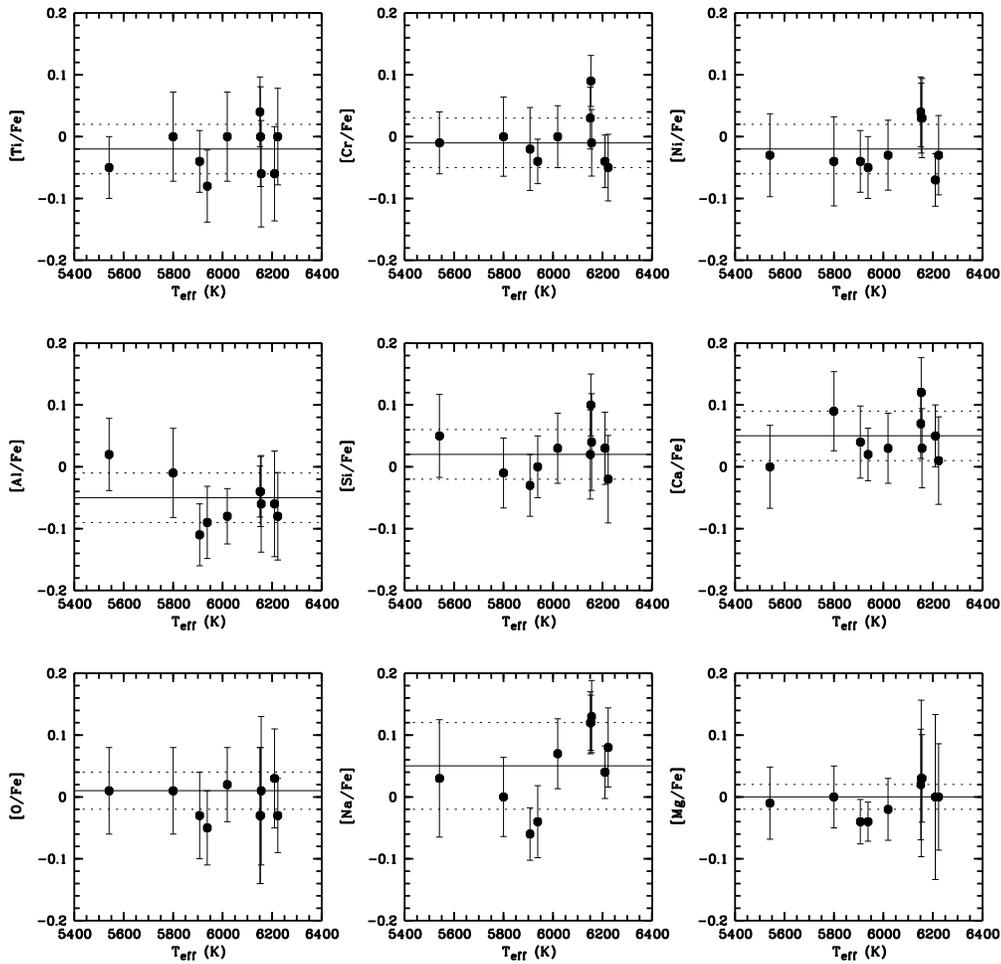, width=14cm} 
\caption{[X/Fe] as a function of effective temperature for the sample
stars. Error bars correspond to the quadratic sum of the rms of [Fe/H]
(or errors due to uncertainties in measured $EW$s for O, whose
abundance has been determined from one line only) and rms of [X/H]. 
As in Fig.~\ref{figure2}, in each panel the solid
and dotted lines represent the mean value $\pm 1$ standard
deviation.}\label{figure3}
\end{figure*}

\begin{figure*}
\psfig{figure=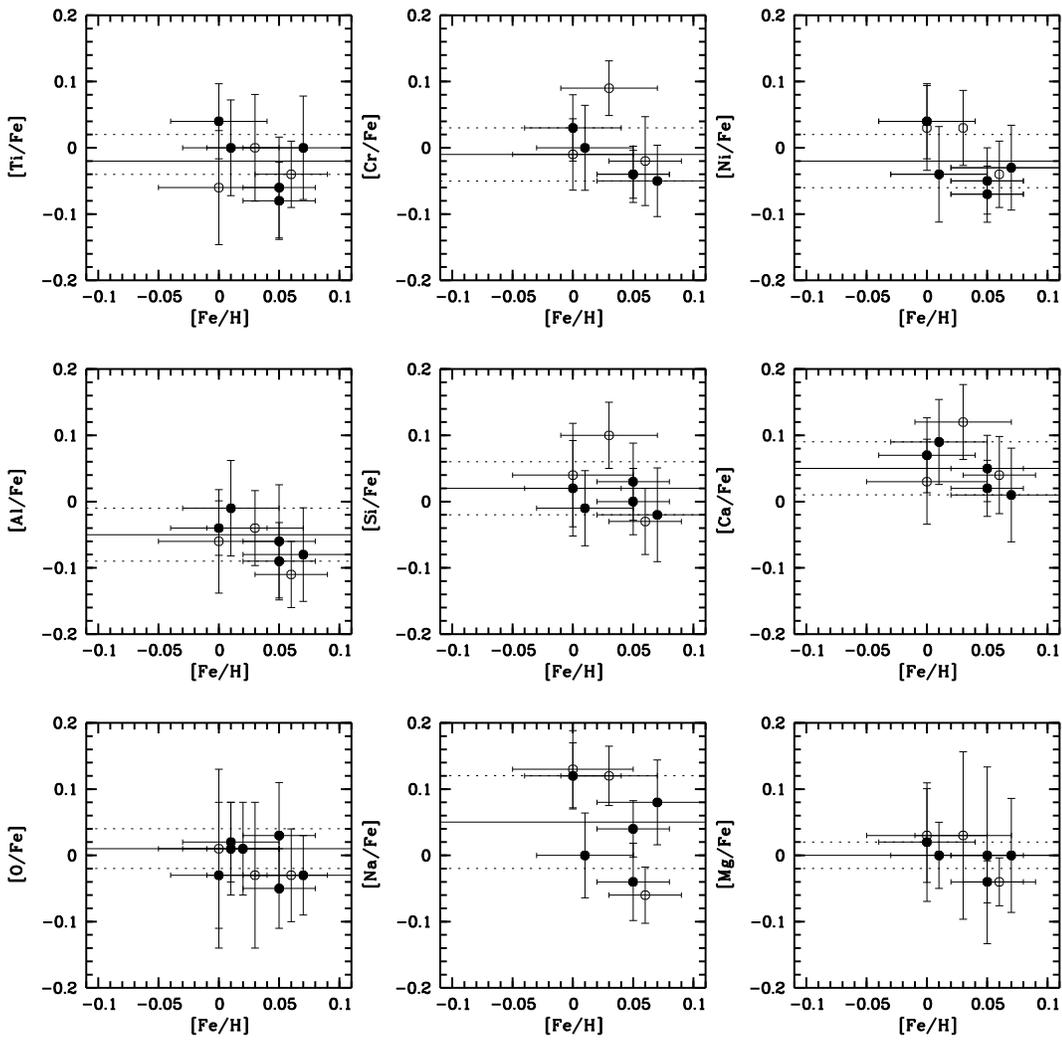, width=15cm, angle=0}
\caption{[X/Fe] vs. [Fe/H] for Li-poor (open circles) and Li-rich
(filled circles) stars. Error bars in [Fe/H] and [X/Fe]
are the same as in Figs.~\ref{figure2} and \ref{figure3}. 
Horizontal lines are the same as in Fig.~\ref{figure3}. The two evolved
stars are not plotted in the figure (see text).
}\label{figure4}
\end{figure*}

Figures~\ref{figure2} and \ref{figure3} together with Tables~\ref{sample},
\ref{ratios}, \ref{oxygen} and \ref{errors} clearly indicate that the standard
deviations $\sigma$ of [Fe/H] and [X/Fe] ratios for the whole sample are 
comparable or even smaller
than measurement uncertainties of individual stars. This on the one
hand implies that
our estimate of internal errors may be somewhat
conservative; on the other hand, and most important,
$\sigma$ values represent upper limits to the dispersion in
[Fe/H] and [X/Fe] ratios, suggesting that 
we can exclude the presence of a scatter in [Fe/H] and [X/Fe]
ratios at a level larger than $\sim 0.05$~dex.
\setcounter{table}{7}
\begin{table*}
\caption{Reanalysis of Tautvai\v sien\. e et al. (\cite{tau00}) data. 
The subscript ``our" indicates the values obtained through our analysis,
while the subscript ``TETI" indicates their original values.}\label{taut_n}
\begin{tabular}{cccccccc}\hline
Ion & \multicolumn{2}{c}{F84} & \multicolumn{2}{c}{F141} & 
\multicolumn{2}{c}{F151} & average$_{\rm our}$\\
 & [X/Fe]$_{\rm TETI}$ & [X/Fe]$_{\rm our}$ & [X/Fe]$_{\rm TETI}$ & [X/Fe]$_{\rm our}$ &[X/Fe]$_{\rm TETI}$ & [X/Fe]$_{\rm our}$ &\\
   & & & & & & &\\
Na~{\sc i} & 0.21 & 0.11 & 0.19 & 0.12 & 0.21 & 0.12 & 0.12\\
Al~{\sc i} & 0.08 & $-0.04$ & 0.12 & $-$0.03 & 0.09 & $-0.01$ & $-$0.03 \\
Si~{\sc i} & 0.09 & $-$0.03 & 0.06 & $-$0.03 & 0.08 & $-0.04$ & $-$0.03 \\ \hline
\end{tabular}
\end{table*}
In order to investigate more in detail the relationship
between heavy element abundances and Li dispersion, 
sample stars shown in Fig.~\ref{figure2} are denoted with different
symbols according to whether they are Li-rich or Li-poor. Analogously,
we show in Fig.~\ref{figure4} [X/Fe] vs. [Fe/H] for our sample
stars with Li-rich and Li-poor objects indicated by different symbols.
We define as Li-rich/poor stars those lying on the upper/lower
envelope of the $\log$~n(Li) vs. \teff~distribution of M~67
(see Pasquini et al. \cite{pas97}; Jones et al. \cite{jon99}). Excluding
stars S1034 and S1239 that are already on the subgiant branch and have
undergone a certain amount of
post-MS Li dilution (Balachandran \cite{bala95}), three
and five of the remaining sample stars fall in the Li-poor/rich category
(open and filled circles, respectively). 
The figure shows that no systematic difference between the
abundance pattern of Li-rich and Li-poor stars is present, in the sense
that Li-poor stars are not systematically more metal-rich/poor and/or have
larger/lower [X/Fe] ratios than the other stars. 
In particular the two stars S1252 and S1256 (similar
temperatures, but a factor of almost 10 difference in Li)
have very similar [Fe/H] and similar (almost identical in some cases)
abundances of $\alpha$-elements, including oxygen, as well as of
Fe-peak elements. As for the other three stars with similar temperatures,
but different Li abundances (S2205 and S988 both Li-poor and S994 Li-rich),
they have similar [X/Fe] ratios for most elements ---again including O---
with the 
exception of Si, Ca, and Cr. [Si/Fe], [Ca/Fe] and [Cr/Fe] 
are somewhat higher in S988 than in the other two stars.
However, not only the difference is within errors, but
the largest difference in heavy
element abundances is seen between the two Li-poor stars themselves,
with S988 being on average more metal rich than S2205. S994 has [X/Fe]
values intermediate between those of S2205 and S988.

We conclude that, based on our sample of M~67 members,
it seems very unlikely that dispersion of a factor about 10 
in Li abundances measured among otherwise similar cluster stars is
due to differences larger than $\sim 0.05$~dex in heavy element abundances.
We note in particular that no dispersion in O -one of the most critical
elements for stellar opacities and thus Li destruction- 
is present among our sample stars.

Our result on
the lack of a dispersion in heavy element abundances leaves
early depletion due to angular momentum loss and transport as the
most likely explanation for the dispersion (Jones et
al. \cite{jon99}). In this scenario, stars with different initial
rotational velocities would undergo different amounts of
Li depletion, resulting in a dispersion in Li at the age of M~67. 
This scenario however encounters two main difficulties:
first, models including mixing due to angular momentum transport
predict a correlated Li and Be depletion, which is instead not
seen in M~67 (Randich et al. \cite{R02}). Second, since observations
of young cluster show that they all have similar rotation distributions,
one would expect that all old clusters should be characterized
by a spread in Li, which is instead not the case.
Further studies are most obviously warranted.
\subsection{M~67 in the disk}
\begin{figure*}
\psfig{figure=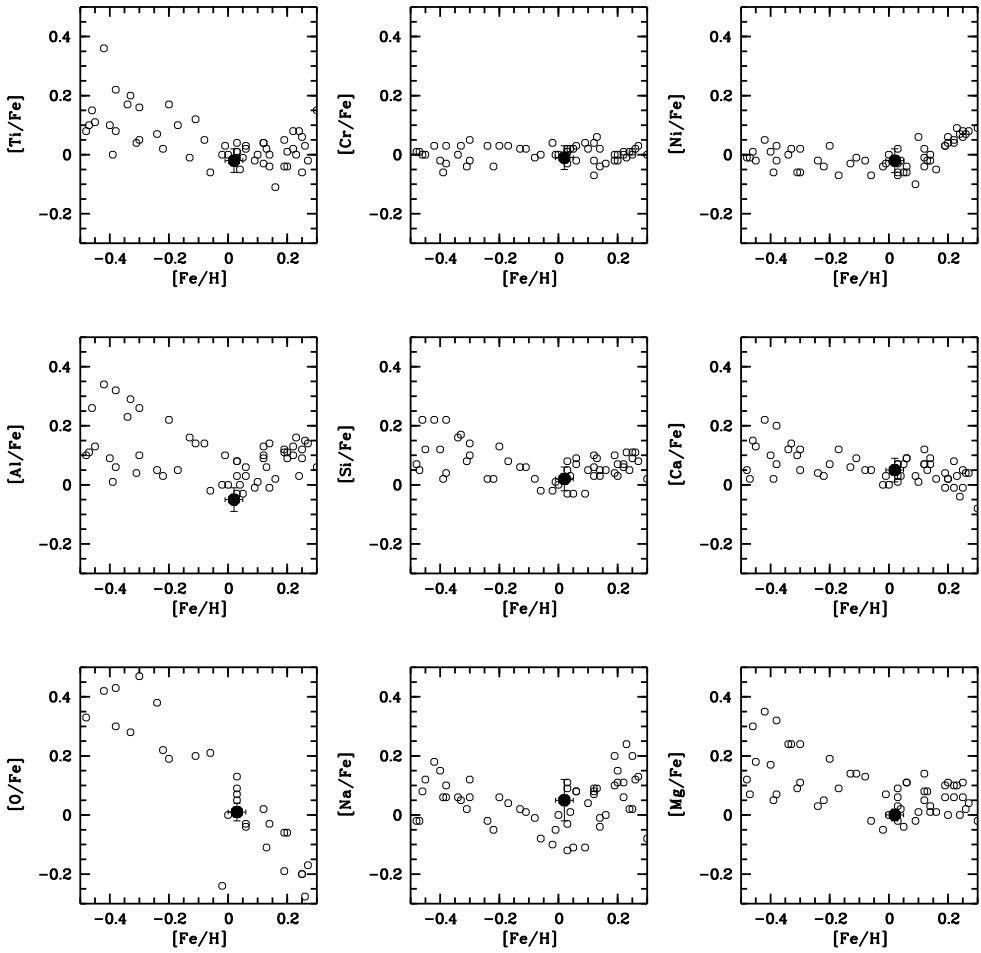, width=14cm}
\caption{[X/Fe] vs. [Fe/H] for field stars from 
Bensby et al. (\cite{bfl03}, \cite{bensby04} ---open
circles)
and the average values from this study (filled symbols).}\label{disk}
\end{figure*}

In Fig.~\ref{disk} we compare the average [X/Fe] vs. [Fe/H] pattern
of M~67 with the sample of field stars
thin and thick disks from Bensby et al. (\cite{bfl03}) and Bensby et al. 
(\cite{bensby04}) for O.
With the caveat that the comparison of M~67 with field stars
can be made only on qualitative grounds, since
abundances are on different scales and systematic offsets might be present,
the figure indicates that the average [X/Fe]
ratios for M~67 very well fit into the general trend of field stars.
The $\alpha$-elements have solar ratios (or slightly below solar
in the case of Ti) and none of 
them is significant over- or under-abundant
with respect to field stars. [O/Fe] is also
solar and very well
fits within the distribution of field stars. Na,
which we find to be somewhat enhanced with respect to the Sun,
lies on the upper envelope of
the distribution of field stars, but is still consistent with it. Similarly,
Al is slightly below the solar ratio and the lower envelope of field
stars, but, given the uncertainties, consistent with it.
In other words, our measurements confirm that M~67 has a ``normal" abundance
pattern for the solar neighborhood and it is very much similar to the 
Sun. As a final remark, we wish to stress that, given its age and 
global abundance
pattern, M~67 represents one of the most promising clusters where to look
for true solar-analogs.

\begin{acknowledgements}
This research has made use of the SIMBAD data base, operated at CDS,
Strasbourg, France. We thank the anonymous referee for the very useful
suggestion of including analysis of oxygen in the present study.
SR and PS acknowledge support by COFIN
2003-029437 by the Italian MIUR.
\end{acknowledgements}

{}
\clearpage
\onecolumn
\Online
\setcounter{table}{1}
\begin{longtable}{lcccr}
\caption{Line list, adopted $\log$~gf values, and solar EWs.
(1): VALD; (2): Lambert \& Walner (\cite{lw}); 
(3): Chen et al. (\cite{chen});  (4): Bensby et al. (\cite{bfl03});
(5): Clementini et al. (\cite{clem}); (6): Fulbright (\cite{full});
(7): Bi\'emont et al (\cite{bie91});
(*): inverse solar analysis.}\label{tab_gf}\\
\hline
Element & $\lambda$~(\AA) & $\log$~gf & ref. &  EW$_{\odot}$ (m\AA)\\
\hline
\endfirsthead
\caption{continued.}\\
\hline
Element    & $\lambda$~(\AA) & $\log$~gf & ref. &  EW$_{\odot}$ (m\AA)\\
\hline
\endhead
\hline\endfoot
O~{\sc i}  & 6300.633 & $-$9.717 & (1) &  5.5\\
Na~{\sc i} & 5682.633 & $-$0.700 & (1) &  107.0\\
Na~{\sc i} & 5688.220 & $-$0.400 & (2) &  130.0\\
Na~{\sc i} & 6154.226 & $-$1.570 & (2) &  35.0\\
Na~{\sc i} & 6160.747 & $-$1.270 & (2) &  52.9\\
Mg~{\sc i} & 5528.405 & $-$0.620 & (1) &  233.0\\
Mg~{\sc i} & 5711.090 & $-$1.724 & (3) &  105.0\\
Al~{\sc i} & 5557.070 & $-$2.160 & (*) &   11.5\\
Al~{\sc i} & 6696.023 & $-$1.540 & (*) &  39.3\\
Al~{\sc i} & 6698.673 & $-$1.890 & (*) &  20.1\\
Si~{\sc i} & 5701.104 & $-$2.050 & (1) &  37.2\\
Si~{\sc i} & 5948.541 & $-$1.230 & (1) &  81.5\\
Si~{\sc i} & 6091.919 & $-$1.400 & (1) &  30.6\\
Si~{\sc i} & 6125.021 & $-$1.550 & (*) &  30.9\\
Si~{\sc i} & 6142.483 & $-$1.480 & (6) &  33.2\\
Si~{\sc i} & 6145.016 & $-$1.440 & (5) &  37.5\\
Si~{\sc i} & 6414.980 & $-$1.100 & (1) &  46.0\\
Si~{\sc i} & 6518.733 & $-$1.500 & (1) &  20.5\\
Si~{\sc i} & 6555.463 & $-$1.000 & (1) &  43.6\\
Ca~{\sc i} & 5512.980 & $-$0.480 & (*) &  87.0\\
Ca~{\sc i} & 5581.965 & $-$0.671 & (3) &  94.3\\
Ca~{\sc i} & 5601.277 & $-$0.523 & (3) &  105.5\\
Ca~{\sc i} & 5867.562 & $-$1.610 & (*) &  24.4\\
Ca~{\sc i} & 6102.723 & $-$0.862 & (1) & 126.5\\
Ca~{\sc i} & 6122.217 & $-$0.386 & (1) & 171.2\\
Ca~{\sc i} & 6161.297 & $-$1.293 & (1) & 60.5\\
Ca~{\sc i} & 6166.439 & $-$1.156 & (1) & 68.6\\
Ca~{\sc i} & 6169.042 & $-$0.804 & (1) & 87.0\\
Ca~{\sc i} & 6169.563 & $-$0.527 & (1) & 107.6\\
Ca~{\sc i} & 6455.598 & $-$1.400 & (*) & 56.6\\
Ca~{\sc i} & 6499.650 & $-$0.818 & (3) & 87.0\\
Ti~{\sc i} & 4805.415 & 0.150 & (1) & 32.6\\
Ti~{\sc i} & 4820.411 & $-$0.441 & (1) & 44.9\\
Ti~{\sc i} & 4885.079 & 0.358 & (1) & 64.0\\
Ti~{\sc i} & 4913.614 & 0.160 & (1) & 53.7\\
Ti~{\sc i} & 5016.161 & $-$0.574 & (1) & 67.5\\
Ti~{\sc i} & 5219.702 & $-$2.292 & (1) & 28.1\\
Ti~{\sc i} & 5866.451 & $-$0.840 & (1) & 48.1\\
Ti~{\sc i} & 5953.160 & $-$0.329 & (1) & 33.6\\
Ti~{\sc i} & 5965.828 & $-$0.409 & (1) & 29.9\\
Ti~{\sc i} & 6258.102 & $-$0.431 & (3) & 50.9\\
Ti~{\sc i} & 6261.098 & $-$0.479 & (1) & 49.8\\
Cr~{\sc i} & 4936.335 & $-$0.340 & (1) & 42.0\\
Cr~{\sc i} & 5247.566 & $-$1.640 & (1) & 81.2\\
Cr~{\sc i} & 5296.691 & $-$1.400 & (1) & 92.2\\
Cr~{\sc i} & 5300.744 & $-$2.120 & (1) & 59.8\\
Cr~{\sc i} & 5329.142 & $-$0.064 & (1) & 66.0\\
Cr~{\sc i} & 5348.312 & $-$1.290 & (1) & 96.9\\
Fe~{\sc i} & 4835.868 & $-$1.500 & (1) & 46.6\\
Fe~{\sc i} & 4875.878 & $-$2.020 & (1) & 55.6\\
Fe~{\sc i} & 4907.732 & $-$1.840 & (1) & 58.7\\
Fe~{\sc i} & 4999.113 & $-$1.740 & (1) & 31.2\\
Fe~{\sc i} & 5036.922 & $-$3.068 & (1) & 22.2\\ 
Fe~{\sc i} & 5044.211 & $-$2.038 & (1) & 75.1\\
Fe~{\sc i} & 5067.150 & $-$0.970 & (1) & 70.0\\
Fe~{\sc i} & 5141.739 & $-$2.190 & (1) & 88.8\\
Fe~{\sc i} & 5162.273 & 0.020 & (1) & 138.1\\
Fe~{\sc i} & 5217.389 & $-$1.070 & (1) & 110.9\\
Fe~{\sc i} & 5228.377 & $-$1.290 & (1) & 52.1\\
Fe~{\sc i} & 5285.129 & $-$1.640 & (1) & 26.7\\
Fe~{\sc i} & 5293.959 & $-$1.870 & (1) & 29.0\\
Fe~{\sc i} & 5373.709 & $-$0.860 & (1) & 63.0\\
Fe~{\sc i} & 5386.334 & $-$1.770 & (1) & 32.0\\
Fe~{\sc i} & 5397.618 & $-$2.480 & (1) & 23.1\\
Fe~{\sc i} & 5472.709 & $-$1.495 & (1) & 42.8\\
Fe~{\sc i} & 5522.447 & $-$1.550 & (1) & 41.4\\
Fe~{\sc i} & 5539.280 & $-$2.660 & (1) & 16.2\\
Fe~{\sc i} & 5543.150 & $-$1.570 & (1) & 63.3\\
Fe~{\sc i} & 5543.936 & $-$1.140 & (1) & 60.2\\
Fe~{\sc i} & 5546.992 & $-$1.910 & (1) & 23.9\\
Fe~{\sc i} & 5584.765 & $-$2.320 & (1) & 34.8\\
Fe~{\sc i} & 5636.696 & $-$2.610 & (1) & 19.1\\
Fe~{\sc i} & 5638.262 & $-$0.870 & (1) & 73.8\\
Fe~{\sc i} & 5662.516 & $-$0.573 & (1) & 93.4\\
Fe~{\sc i} & 5691.497 & $-$1.520 & (1) & 39.6\\
Fe~{\sc i} & 5701.545 & $-$2.216 & (1) & 84.8\\
Fe~{\sc i} & 5862.353 & $-$0.058 & (1) & 86.4\\
Fe~{\sc i} & 5916.247 & $-$2.994 & (1) & 54.2\\
Fe~{\sc i} & 5930.180 & $-$0.230 & (1) & 88.8\\
Fe~{\sc i} & 5934.655 & $-$1.170 & (1) & 72.6\\
Fe~{\sc i} & 5956.694 & $-$4.605 & (1) & 54.1\\
Fe~{\sc i} & 5976.775 & $-$1.310 & (1) & 66.7\\
Fe~{\sc i} & 5984.814 & $-$0.343 & (1) & 82.7\\
Fe~{\sc i} & 5987.066 & $-$0.556 & (1) & 66.8\\
Fe~{\sc i} & 6003.012 & $-$1.120 & (1) & 77.8\\
Fe~{\sc i} & 6024.058 & $-$0.120 & (1) & 108.7\\
Fe~{\sc i} & 6056.005 & $-$0.460 & (1) & 72.6\\
Fe~{\sc i} & 6078.491 & $-$0.424 & (1) & 76.0\\
Fe~{\sc i} & 6136.995 & $-$2.950 & (1) & 68.6\\
Fe~{\sc i} & 6157.728 & $-$1.260 & (1) & 61.8\\
Fe~{\sc i} & 6187.990 & $-$1.720 & (1) & 46.2\\
Fe~{\sc i} & 6200.313 & $-$2.437 & (1) & 73.5\\
Fe~{\sc i} & 6315.811 & $-$1.710 & (1) & 40.3\\
Fe~{\sc i} & 6322.685 & $-$2.426 & (1) & 75.0\\
Fe~{\sc i} & 6336.824 & $-$0.856 & (1) & 105.5\\
Fe~{\sc i} & 6344.149 & $-$2.923 & (1) & 61.0\\
Fe~{\sc i} & 6469.193 & $-$0.770 & (1) & 54.8\\
Fe~{\sc i} & 6495.742 & $-$0.940 & (1) & 42.2\\
Fe~{\sc i} & 6498.939 & $-$4.699 & (1) & 46.6\\
Fe~{\sc i} & 6574.228 & $-$5.023 & (1) & 29.4\\ 
Fe~{\sc i} & 6609.110 & $-$2.692 & (1) & 66.2\\
Fe~{\sc i} & 6703.567 & $-$3.160 & (1) & 38.1\\
Fe~{\sc i} & 6733.151 & $-$1.580 & (1) & 23.1\\
Fe~{\sc i} & 6750.153 & $-$2.621 & (1) & 74.2\\
Fe~{\sc i} & 6806.845 & $-$3.210 & (1) & 33.7\\ 
Fe~{\sc ii}& 5264.812 & $-$3.120 & (1) & 47.1\\
Fe~{\sc ii}& 5325.553 & $-$3.222 & (7) & 41.1\\
Fe~{\sc ii}& 5414.073 & $-$3.750 & (7) & 23.7\\
Fe~{\sc ii}& 5425.257 & $-$3.372 & (7) & 37.8\\
Fe~{\sc ii}& 5991.376 & $-$3.557 & (7) & 31.4\\
Fe~{\sc ii}& 6084.111 & $-$3.808 & (7) & 21.5\\
Fe~{\sc ii}& 6149.258 & $-$2.724 & (7) & 34.8\\
Fe~{\sc ii}& 6247.557 & $-$2.329 & (7) & 53.6\\
Fe~{\sc ii}& 6432.680 & $-$3.708 & (7) & 39.4\\
Fe~{\sc ii}& 6456.383 & $-$2.075 & (7) & 64.4\\
Fe~{\sc ii}& 6516.080 & $-$3.450 & (7) & 49.3\\
Ni~{\sc i} & 4806.984 & $-$0.640 & (1) & 59.2\\
Ni~{\sc i} & 4852.547 & $-$1.070 & (1) & 42.7\\
Ni~{\sc i} & 4904.407 & $-$0.170 & (1) & 87.9\\
Ni~{\sc i} & 4913.968 & $-$0.630 & (1) & 58.7\\
Ni~{\sc i} & 4946.029 & $-$1.290 & (1) & 23.5\\
Ni~{\sc i} & 5003.734 & $-$3.130 & (4) & 32.8\\
Ni~{\sc i} & 5010.934 & $-$0.870 & (1) & 52.2\\
Ni~{\sc i} & 5032.723 & $-$1.270 & (1) & 20.6\\
Ni~{\sc i} & 5082.339 & $-$0.590 & (*) & 62.8\\
Ni~{\sc i} & 5155.125 & $-$0.650 & (1) & 47.9\\
Ni~{\sc i} & 5435.855 & $-$2.590 & (1) & 45.8\\
Ni~{\sc i} & 5462.485 & $-$0.930 & (1) & 39.9\\
Ni~{\sc i} & 5589.357 & $-$1.140 & (1) & 27.1\\
Ni~{\sc i} & 5593.733 & $-$0.840 & (1) & 41.6\\
Ni~{\sc i} & 5625.312 & $-$0.700 & (1) & 38.6\\
Ni~{\sc i} & 5641.880 & $-$1.070 & (1) & 24.1\\
Ni~{\sc i} & 5682.198 & $-$0.499 & (3) & 50.1\\
Ni~{\sc i} & 6111.066 & $-$0.870 & (1) & 33.4\\
Ni~{\sc i} & 6175.360 & $-$0.559 & (1) & 48.4\\
Ni~{\sc i} & 6186.709 & $-$0.960 & (1) & 29.0\\
Ni~{\sc i} & 6191.171 & $-$2.353 & (1) & 74.9\\
Ni~{\sc i} & 6223.981 & $-$0.910 & (1) & 26.2\\
Ni~{\sc i} & 6378.247 & $-$0.830 & (1) & 33.3\\
Ni~{\sc i} & 6586.308 & $-$2.810 & (1) & 41.0\\
\end{longtable}
\end{document}